\documentclass[12pt,draftcls, peerreview, onecolumn]{IEEEtran}
\usepackage{url}
\usepackage{cite}      

\usepackage{graphicx}  
\usepackage{epsf}
\usepackage{epsfig}

\usepackage{subfigure} 

\usepackage{url}       

\usepackage{amsmath}   
\usepackage{xcolor}

\usepackage{comment}

\makeatletter \newcommand{\vast}{\bBigg@{3}} \newcommand{\vvast}{\bBigg@{4}} \newcommand{\Vast}{\bBigg@{5}} \makeatother 
\DeclareMathAlphabet{\mathcalligra}{T1}{calligra}{m}{n}

\renewcommand{\baselinestretch}{2}

\begin{document}

\title{
{Development of an Ultrahigh Bandwidth Software-defined Radio Platform}}

\author{
Sung~Sik~Nam, Changseok Yoon, Ki-Hong Park
	and Mohamed-Slim~Alouini
\thanks{
S.~S.~Nam is in Gachon University, Korea (Email : ssnam@gachon.ac.kr).  C.~S. Yoon is in Korea Electronics Technology Institute, Korea. K.-H. Park and M.-S. Alouini are with the Computer, Electrical and Mathematical Science and Engineering Division (CEMSE), King Abdullah University of Science and Technology (KAUST), Thuwal, Saudi Arabia (Email: \{kihong.park, slim.alouini\}@kaust.edu.sa).}}

\maketitle
\vspace{-0.8cm}
\begin{abstract}
For the development of new digital signal processing systems and services, the rapid, easy, and convenient prototyping of ideas and the rapid time-to-market of products are becoming important with advances in technology.
Conventionally, for the development stage, particularly when confirming the feasibility or performance of a new system or service, an idea is first confirmed through a computer-based software simulation after developing an accurate model of the operating environment. Next, this idea is validated and tested in the real operating environment. 
The new systems or services and their operating environments are becoming increasingly complicated. Hence, their development processes too are more complex cost- and time-intensive tasks that require engineers with skill and professional knowledge/experience. Furthermore, for ensuring fast time-to-market, all the development processes encompassing the (i) algorithm development, (ii) product prototyping, and (iii) final product development, must be closely linked such that they can be quickly completed.
In this context, the aim of this paper is to propose an ultrahigh bandwidth software-defined radio platform that can prototype a quasi-real-time operating system without a developer having sophisticated hardware/software expertise. This platform allows the realization of a software-implemented digital signal processing system in minimal time with minimal efforts and without the need of a host computer.
\end{abstract}

\vspace{-0.8cm}


\clearpage
\tableofcontents
\clearpage

\section{Introduction}  \label{intro}
When developing new algorithms for communication and sensing applications, real-time operational testing is often necessary for analyzing the performance of the final product and during development, and testing is performed during development right from the initial algorithm development stage to the final application testing stage.

The commonly used software-based simulation tests and/or demonstrations based on expensive equipment do not have real-time testing capabilities that are important to confirm the performance of real end products. Hence, only limited tasks can be validated. 
To perform an effective test, it is necessary to analyze and understand the system of target applications, implement complex hardware design based on this system, and build the hardware that can test the algorithm.

Understanding this complete process is not always necessary for researchers and engineers who require to focus on algorithm development. Building the hardware requires considerably more time and effort than that required to develop the actual algorithm. This project aims to reduce the wastage of time and effort because of unnecessary hardware implementation in the development of algorithms, facilitate real-time testing of developed algorithms, and develop a system that simplifies the processes to the final stage of development.

In general, when designing a digital signal processing (DSP) system (the most important process in system development), we develop an algorithm, system, or service from new ideas based on a specific target model. 
After confirming the performance of the proposed algorithm, system, and related services through personal computer (PC)-based software simulation (e.g., using standard programs such as MATLAB, Python, and Octave), an environment model is developed and implemented by reflecting the characteristics of the actual operating environment. Then, the proposed idea should be confirmed and tested based on this modeled operating environment.

For simple systems, software simulations can confirm the proposed idea at a certain level; however, systems or services and their operating environments have become increasingly complex in recent years.
In such an environment, after confirming the effectiveness or performance of a new system or service through software simulation, there is a limitation for solving multiple problems that may arise during application to an actual system in a real environment.
Rather than using a PC-based simulation based is a modeled operating environment, it is more meaningful and effective to use a hardware-based prototype product in a real environment.
Therefore, to overcome these limitations, this idea must be applied and tested on a hardware-based real-time prototype product after validation via simulation.

Skilled hardware expertise is required to develop an idea testing system based on an actual operating system. Moreover, even an engineer with experienced professional expertise requires considerable time and effort to solve multiple problems that may arise during the experiment and hardware development processes.
This results in a significant increase in development time.
Furthermore, as per the latest technology trends, the latest systems require ultrahigh bandwidth ($>1$ GHz) and high-speed operation as important elements.

In this context, our aim is to develop a programmable and reconfigurable DSP platform for wireless applications with ultrahigh bandwidth ($>1$GHz) that can operate stand-alone. Using this proposed single platform, we can prototype a stand-alone real-time operating system without a host PC connection, extremely expensive test equipment, or a skillful developer with sophisticated hardware/software expertise.
This platform will enable developers to quickly realize a software-implemented system and will meet the consumers' desire for new technologies and services against the background of the rapidly changing market and technology trends in a timely manner. 
In particular, our aim is to propose a next-generation ultrahigh bandwidth software-defined radio (SDR) platform capable of DSP in the GHz range. This platform will include ultra-fast data converters, e.g., analog-to-digital/digital-to-analog converters (ADCs/DACs), field programmable gate arrays (FPGAs), and microcontroller units (MCUs), and comprises hardware logic, firmware and application software to control them. This platform will enable the effective implementation of DSP stages for multiple high-speed communication and sensing applications such as free-space optical communication (FSO), high-speed underwater optical communication (UWOC), millimeter wave communication, terahertz (THz) communication systems, and light detection and ranging (LIDAR).

\section{Current trends of communication system technology and the limitations of implementing it with the existing conventional SDR platform} \label{sec_1}

\subsection{Current Trends of Technology}
Recently, in communication technology, particularly for wireless networks, the research and development for a communication system requiring ultrahigh bandwidth has gained considerable interest and an unprecedented increase in the on-demand traffic. The upcoming wireless network will require ultrahigh data exchange in terms of (1) mobile broadband traffic and extreme capacity Xhaul-supporting live multimedia streaming, multi-sensory extended reality (XR), and holographic 3D conferencing; (2) ultra-low latency and high reliability in highly autonomous systems that jointly control communication, computing, control, localization and sensing (3CLS), e.g., connected robotics and aerial vehicles; (3) massive connectivity, scalability, and heterogeneity, e.g., the Internet of everything (IoE) and Industrial 4.0; (4) wireless intelligence enhancing edge computing and semantic communications; and (5) security in blockchain and distributed ledger technology\cite{6824746}.

A high level of network flexibility is necessary to efficiently use the network resources to satisfy service-specific data on demand in a heterogeneous network environment. Network function virtualization (NFV) guarantees additional flexibility to expand the network capabilities by decoupling the software implementation of the network functions from the traditional hardware appliances. 
Virtualization technologies and commercially available reconfigurable/programmable hardware can be leveraged to facilitate the agile deployment of new network services and can ultimately make the deployment cost-effective by reducing capital expenditure (CAPEX) and operating expenditure (OPEX)\cite{7045396}.

Therefore, we have seen considerable interest in programmable and reconfigurable hardware platforms such as SDR that supports the ultrahigh speed DSP capability required in ultrahigh bandwidth communication systems such as 5G cellular, mmWave, and FSO communications and in ultra-precision sensor/localization systems.
The use of ultrahigh bandwidth signals is expanding to 
increase the data transfer rate for communication systems and achieve higher precision for sensor systems. As per recent technology trends, ultrahigh bandwidth ($>1$ GHz) and high speed operation are the key elements.
However, the implementation of such a high-speed system, particular considering real-time operations is very complex and results in a significant increase in the development complexity.

\subsection{Limitations of the Conventional SDR Platforms}
The development of a new ultrahigh bandwidth communication and signal processing system often requires high-performance test equipment or dedicated hardware systems. In terms of high-performance test equipment, the system can typically be configured with a digital signal analyzer (DSA) for the ADC and an arbitrary waveform generator (AWG) for the DAC connected to a host computer\cite{Oubei:15}. 
These systems can provide high-precision and high-speed sampling but they are usually bulky and expensive. In terms of the latter, developing and confirming a custom dedicated hardware system for the target system or algorithm will eventually lead to increased costs and delayed time to market.
Multiple conventional SDR platforms are available in the market to simplify equipment selection and integration and to quickly design systems using softwares\cite{7378428}.
However, the operating bandwidth of these products is insufficient (e.g., limited to 200 MHz) to support next-generation applications such as 5G New Radio, mmWave, FSO, and LIDAR. 
Furthermore, the existing SDR platform is not designed to operate in a stand-alone mode; it has to be connected to a host computer. 
Such systems usually do not have a real-time/application processor for stand-alone operation. Eventually, the users of these systems will face a lot of trouble to either design real-time applications or provide the mobility in the entire system.
Furthermore, regardless of the target application, the conventional SDR hardware board has a fixed hardware structure. 
Moreover, regardless of the bandwidth and performance requirements, the conventional hardware boards use the same transceiver front end or processing unit, thereby resulting in a higher cost considering the performance required to implement the target system or algorithm or relatively unnecessary power consumption.

\subsection{Limitation of Some Selected Examples of Competitive Platforms under Development}
The ultrahigh bandwidth SDR platform can be used for developing various applications.
In particular, the ultrahigh bandwidth SDR platform has various functions as a stand-alone system, enables simple and rapid development of target systems, and allows the rapid improvement of system/algorithms via interactive testing and verification during development.
Therefore, the ultrahigh bandwidth SDR platform can be used not only by system development professionals, but also by students for learning and development in academic fields.
In the industry, the platform can be used for prototyping the development of new communication and signal-processing systems and thereby can be used for testing various tests and system certifications. 
Furthermore, because the platform can be independently operated, it can be configured to develop a base station for next-generation communication. Moreover, the platform is compact, which allows it to be mounted and operated on mobile vehicles.

An example of applications of ultrahigh bandwidth SDR platforms is the recent research project on a futuristic wireless network platform. The latest flagship project in the United States (US), namely, Platforms for Advanced Wireless Research (PAWR), initiated  the experimental exploration of new wireless devices, communication techniques, networks, systems, and services that will revolutionize the wireless ecosystem in the US while sustaining US leadership and economic competitiveness for many decades. 
This project includes multiple sub-projects on platforms, e.g., Cloud-enhanced Open Software-defined Mobile Test-bed (COSMOS), the Platform for Open Wireless Data-driven Experimental Research-reconfigurable Eco-system for Next-generation End-to-end Wireless (POWDER-RENEW), and Aerial Experimentation and Research Platform for Advanced Wireless (AERPAW).

The COSMOS project plans to build an open software-defined mobile test-bed at an urban level by networking the system connecting universal software radio peripheral (USRP) SDR and Intel computers. Individual devices are extremely large and consume a lot of power. 
Moreover, a large number of wire connections are required for a broadband connection, leading to operation errors. For a city-scale test-bed as intended for the COSMOS, different types of SDR platforms operating even in the same frequency bands are used to design the wireless nodes that may provide unbalanced performance optimization among different SDRs during development or may necessitate additional effort to be inputted because of the individual development process\cite{10.1145/3372224.3380891}.

The POWDER project plans to build an open wireless test platform using an USRP SDR and a custom-made platform (RENEW). 
The RENEW platform has custom-made 64-antenna hardware and RENEW software. The RENEW platform provides MATLAB design flow, real-time measurement flow, and test and analysis design flow. Although the design is extremely good, the overall design process depends on the MATLAB functionalities such as the HDL coder and embedded coder.; furthermore, it is difficult for beginners. 
Moreover. although hardware design is excellent, it is developed only for 64 massive multiple-input multiple-output (MIMO) systems comprising daisy-chained SDR devices. Hence, it cannot be used for general purposes. An open-source generalized API called SoapySDR is used to interface, configure, and stream with an individual SDR module. It offers the advantages of platform-independent interfacing and open-source applications for hardware support. However, it is not always specified to optimize the configuration that fots the high performance of specific SDR platforms, and hence the users who develop and integrate the algorithm in SDR platform have to expend more effort, time, and labor to learn this API.

The AERPAW Project plans to build aerial experimentation and research platforms using the USRP SDR equipped with an additionally developed add-on board. Furthermore, to control this, an additional general-purpose PC must be installed. Such a system is bulky, and high power consumption, which currently limits the flight time of aerial vehicles to 20-40 min. This imposes a requirement for high payload performance for individual aerial vehicles, and the vehicles require aditional batteries\cite{9061157}.

These platforms cannot provide a total solution for turning ideas into final products. In general, various equipment and tools are required in the development stages. Each stage has a different work process and knowledge requirements. Development stages are sometimes disconnected regardless of each other. Therefore, the nature of the development process, which generally proceeds in several stages, makes flexible overall development difficult.

For seamless interworking between digital simulation models and real operating systems, the ultrahigh bandwidth SDR system provides effective design methods, interfaces, and operation methods. 
Such interworking eventually enables successful development via continuous feedback between the simulation model and implemented system even for system development.
It helps monitor and debug the condition of the product even in real operating environments. To summarize, this proposed SDR can be effectively used in all phases of design, development, and verification. 
Moreover, our products can provide a small size, weight, and power (SWaP) wireless communication edge computing hardware platform for various applications, including unmanned aerial vehicle (UAV) applications.

\begin{figure}
\centering
\includegraphics[width=6.5in]{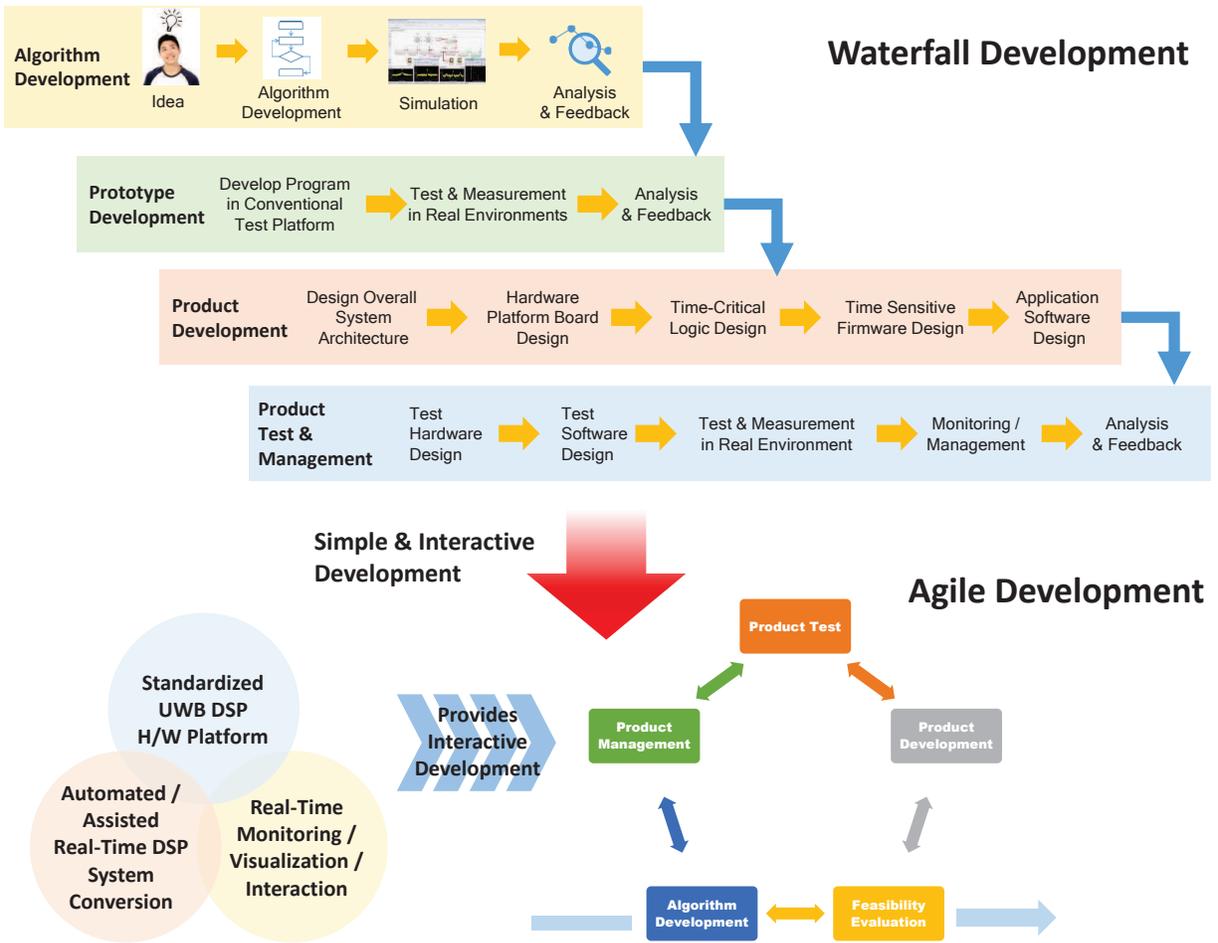}
\caption{Comparison of development process between conventional and proposed platforms.}
\label{Fig.1}
\end{figure}

\section{UltraHigh Bandwidth SDR platform} \label{sec_2}
As shown in Fig.~\ref{Fig.1}, conventional software-based DSP system development is based on the legacy waterfall type development method. The product life cycle basically comprises algorithm development, verification of feasibility through prototyping, actual development, and testing and management. 
Because the DSP system includes hardware, whenever development proceeds stepwise, the transition between steps cannot smoothly proceed because the development proceeds based on the environment built in the previous step. 

The proposed ultrahigh bandwidth SDR platform, i.e., the UltraSDR platform enables movement between each process during development. In recent years, the agile development methodology provides this development freedom for platforms and has been predominantly used in software development. When implementing a new idea, agile development makes it possible to fluidly and smoothly move between the development stages, such that development can smoothly proceed. Therefore, the biggest advantage of the proposed platform is that it can efficiently obtain high-quality results in a shorter time than the time taken by the conventional platform.

The detailed technical requirements and impacts of the proposed UltraSDR platform are described in the following subsection.

\subsection{Technical Requirements of the Proposed UltraSDR Platform to Overcome the Limitations of the Existing Platforms}

\subsubsection{User-centric Development Platform}
Compared to conventional platforms, the ultrahigh bandwidth SDR platform should place emphasis  on user convenience.
In general, by accurately understanding the hardware and spftware specifications of the platform used for real-time system configuration, most existing SDR platforms require proper control for real-time operation. During system development, the user must manually set and control the platform, and any operational problems that arise are the responsibility of the user. Hence, it is important for users to understand accurately not only the representative operation but also the trivial operation of the hardware platform. 
This becomes a big barrier for engineers other than hardware experts when developing a prototype or commercial system based on the SDR platform. 
Therefore, considering the basic purpose of the SDR platform, there is a requirement for a platform that allows general users, in addition to  hardware experts, to develop a system that  performs the desired operation of target applications accurately without knowledge of the hardware operation within the SDR platform. The proposed platform allows users to generate and analyze signals, and automatically configure a real-time operating platform simply by describing the requirements for the real-time operation of the target system. Thus, the user can focus on the actual development and testing of the core algorithm.

\subsubsection{Standardized Hardware Platform}
Effective and stable system development is made possible by applying a standardized hardware platform for system development.
In the existing commercial platforms, the software and hardware development processes are separated to enable operation in different hardware during the system development process. This separation makes it difficult for users to develop an interface with the target hardware platform.
This results in many problems related to the interface and, much time is required to solve them. However, because the proposed system develops software based on a standardized hardware platform, these types of problems can be avoided during the development.

\subsubsection{Easy Development Environments}
The proposed platform provides an integrated development tool for real-time design to enable users to focus on core algorithm design.
The existing platforms generally provide platform-specific development languages or tools for the real-time design of systems \cite{7378427}.
Hence, general users and even experts require to go through additional cumbersome and additional learning for development with the existing platforms.
In particular, to develop a real-time system, the system must be so designed as to allow the real-time operation of the system considering the system performance and operation aspects. Therefore, additional time is spent in contemplating this limitation than in developing the core algorithm.
As the proposed system provides a real-time scheduler function to integrate hardware and software, this inconvenience can be reduced from the user's viewpoint.

\subsubsection{Interactive Test and Verification}
The proposed platform provides a system interaction tool that can continuously monitor, analyze, and improve the target system even during system development and operation.
The existing platforms do not provide hardware to monitor the operation and performance of the developed system, making it difficult to capture signals in real-time or perform tasks such as real-time analysis and on-site emulation for various operating situations.
However, the proposed system provides a real-time interconnection manager to connect an actual operating system and a host personal computer (PC), as well as allows users to conveniently monitor, analyze, and improve the target system and in particular, the system performance during development.

\subsection{Impact of the Proposed SDR Platform}
When developing novel algorithms in various application fields, it is possible to use a single platform for development from the initial operation verification to the final application test stage.
Hardware logic, device firmware, and applications can be developed and integrated on a single platform, allowing testing in not only a lab environment but also a real environment wherein the target system will be operated.  Hence, it will be possible to perform the same motion test as that for an actual product.
Because multiple algorithms and applications can be developed and prototyped within a short time, the time required for testbed or product development can be considerably lesser.
A single platform can meet the requirements of various target applications in terms of size, performance, and power consumption.
Therefore, the platform can be effectively applied directly to develop a wide range of applications with a single platform framework.

At present, even after product launch, the over-the-air programming (OTA programming) technology that improves system functions and performance through continuous remote updates is extensively used, and it provides a competitive advantage for products.
The above mentioned functions provided by the proposed UltraSDR platform make it easy to apply the OTA function to the target system. 
For various target systems, the systems developed based on UltraSDR have a standardized system structure.  The developed new functions/results can be collectively applied through the OTA function on multiple and various platforms.
Moreover, a new algorithm can be quickly developed/tested by applying the automatic/design support conversion function to the code applicable for the real-time DSP system, thereby reducing the OTA cycle.
Furthermore, for updating new algorithm, it is possible to collect and analyze data about the algorithm performance in the actual operating environment of the product through the Real-Time Monitoring/Visualization/Interaction function and report these results, thereby enabling the rapid improvement of the algorithm.

\section{Technical Challenges in Implementing the Proposed UltraSDR Platform} \label{sec_3}

\subsection{User Centric Development Platform}
The challenges for designing a user-friendly platform is that the platform should be designed such that users with different hardware and software knowledge levels from the  beginner to expert level can use the platform without difficulty to obtain the final target result. Generally, when the hardware experts encounters a new platform, they can effectively identify the functions of the platform to be used for development based on their experience. Then, they can develop the target product using the new platform in a short time period by learning the parts required to additionally use the platform. However, beginner-level users may not understand the various functions in the platform, and even after understanding the functions, they may find it complicated to understand how to develop the target system using these functions.

The existing SDR platforms use their own software-based framework to automate the entire development process and provide various libraries and demo codes in the framework suitable for various application. 
Therefore, users generally expect the SDR platforms to provide various operation functions rather than only limited operation functions for a specific target\cite{7086409}.
Users adopt the SDR platform considering the advantages in terms of development convenience in that all tasks necessary for development can be automated during the development process. However, for the user to accurately understand and use these functions, it is necessary foor them to thoroughly understand the entire platform from hardware to software; this may be a challenge even for experts. 
Hence, despite the original intent of the SDR platform, beginners may find it extremely tough to use these platforms for system development.

Therefore, a user-centric development platform should first help users get familiarized  with the platform based on the features they already know.
Furthermore, the users will require to master the knowledge of new features and learn to use them stepwise. Finally the users will use the features they need to achieve the end goal result.
In other words, learning about the platform may occur from the user's point of view during the development process, and the final result can be implemented in a stepwise manner.
Hence, an approach that can identify and leverage the necessary functions stepwise and  tools for the same are required.

For example, the first step of development provides the function to test the results of a PC-based simulation code based on sequential execution familiar to general users on the platform. Although this function does not satisfy the detailed requirements such as real-time performance of the target application, the user can verify the design result of the target application.

After confirming this design result, the next step will be to reflect the information necessary for real-time computational processing of the target application (e.g., processing time limit, iterative execution, and conditional computation in the simulation-based code.
Then, as per the results of the preliminary review, a it goes through the steps of gradually transition to an optimized operation is achieved. Finally, not only experts but also beginners can develop a system that independently operates in real time.

\subsection{Standardized Hardware Platform}
The challenging task in designing a standardized platform is to provide the flexibility to implement various applications, while also providing stability to ensure that unexpected operation problems do not occur during development.

In general, each application is developed using a separate optimized platform. However, if these targets are developed using a standardized hardware platform, the required effort, time, and cost will be lesser compared to those when using a separate platform specialized for each application each time. However, if a standardized platform provides various functions, the system may face stability problems such as unexpected operation problems during target system development. Then, users may need to solve the operation problems on the platform,  and thus eventually time and effort are wasted.

The existing SDR platforms provided the flexibility by using the common software framework supporting multiple hardware platforms. This method has the advantage that one software application can be distributed and utilized regardless of the hardware platform. However, since the functions and performance provided by each hardware platform are different, it is not certain whether the contents designed based on the common software framework can work properly for each hardware according to the user's needs. For example, since the iPhone can perform integrated operation verification from hardware to software, it is possible to guarantee the operation performance of the result, but since Android provides only a software platform, the operation performance of the result in each hardware to which it is applied cannot not be guaranteed.

Therefore, a standardized hardware structure is essential to ensure stable operation to minimize unexpected problems when developing various applications. For example, to achieve this purpose, standardized computation units (CPU/FPGA architecture, operating speed, and computation accelerators), peripherals (GPU, H.264/H.265 video encoder/decoder, etc.), and a standardized connection structure for their connection (such as the AXI4 interface between processor and peripheral devices, ADC/DAC module, and JESD204B/C interface) should be provided. In addition, for flexible expansion in terms of the operating performance, there is a need for an expandable connection link that can expand additional functions in parallel as needed.

A standardized SDR platform should determine/include necessary computing devices considering the real-time aspect of computing devices inside the platform and should allow the design to connect them effectively. First, the programmable logic (PL) for processing data at the highest speed/repetition, a real-time processing unit (RPU) for calculations requiring high real-time performance, an application processing unit (APU) for processing various application operations based on the operating system unit, and a graphical processing unit (GPU) for processing multimedia operations should be included. The connection of these standardized computing devices should be accessed through a standardized complex connection structure such as the AXI interface, GPIO interface, or JESD204B/C. In addition, for module-based expansion, there is a need to include a high-speed hardware interface between the modules and a software management interface that can distribute and synthesize operations based on modules.

\subsection{Easy Development Environment}
The challenge in real-time development environment design is that it must be possible to comprehensively plan and realize the design required by the user with the goal of real-time operation by utilizing the multiple processing devices available in the platform. To develop various target applications on one standardized platform, several necessary sub-platforms should be configured heterogeneously. Therefore, for these heterogeneous subplatforms to operate organically in real-time, each operation of the heterogeneous subplatforms should be predicted and complex scheduling tasks should be performed to operate these subplatforms based on the prediction results.
However, since the individual computing units (i.e., the CPU, MCU, and FPGA) inside these sub-platforms operate independently, depending on the target application, it may be easy or very difficult to predict the operation of the individual computing units inside each of these subplatforms.

In the existing SDR platforms, the performance of individual computing devices (i.e., FPGA, CPU, and GPU) inside the platform has been gradually improved. However, data operation and data transfer between these computing devices were not accordingly performed. Consequently, the system operation processing performance is a relatively lower than the actual platform performance. This is because the scheduler of the general operating system is used to increase the degree of freedom. To improve the performance, a customized system capable of optimizing the operation schedule at a low level in units of operating frequencies can be used \cite{9180885}; however, this also increases the difficulty of development thereby leading to a longer development period.

A computational profiling and planning tool that can comprehensively predicts the operation of individual computational devices inside each subplatform in advance is necessary to realize improved operation processing performance and ease to development. Based on the profiling result data, comprehensive planning is performed for the operation time of each computing devices to check whether the operation can be effectively performed. This design result should be monitored and controlled by a dedicated integrated scheduler hardware that manages all individual computing devices.

The calculation profiling tool should be designed based on the initial simulation code to predict the required calculation time when a subplatform is changed.
Also, based on this calculation, it should be possible to change the subplatform on which the unit operation operates as needed during planning. For example, when parallelization of continuous data processing is possible, each operation can be parallelized and implemented in the FPGA, so that subsequent operation processing can be performed according to the output timing of the determined operation result. When the design related to quasi-real-time operation or real-time operation is completed through the planning process, this content is saved as action time and action period in the dedicated scheduler hardware, and a signal for starting operation is transmitted to each subplatform. If the target operation is not completed within the desired time, this content should also be reported to the scheduler and processed so that it can be operated in real time through subsequent processing.

\subsection{Interactive Test and Verification}
The challenge in the interactive test and verification stage is to collect, analyze, and present the data necessary for verification as simply as possible. In general, to analyze the operation of a ``user-developed system'', additional data besides the essential data must be collected. This eventually leads to a large amount of computation to process and analyze the collected data. If the functional blocks for data collection and processing are implemented in the same hardware platform as the ``user-developed system'', the main operation of this system may be affected because of the additional collection and analysis blocks.

The existing SDR platform do not propose/provide additional probing methods for development. Therefore, in general, the SDR platform itself can only pass data and process the passed data. Therefore, in the past, users had to develop and use monitoring methods themselves to monitor the desired data based on these functions.
For example, commonly used functions (e.g., spectrum analyzer, and scope sink) are provided in the form of a library and can be effectively utilized. However, since these standard verification functions alone cannot provide an appropriate solution to verify the desired part of the system effectively during development, the user must spend additional effort and time to implement a monitoring tool for such verification.

Therefore, the platform should be designed to provide a tool integrating software and hardware, and this tool should be capable of comprehensive monitoring and analysis. The interactive test and verification function should provide a standardized data probing interface separately from the system. The user must be provided core analysis tools for this function. The module in charge of this test and verification function consists of hardware intellectual properties (IPs) that can be defined based on software that operate independently from the development system inside the platform or such a module should be provided as an independent module that can be attached outside the platform.

\section{Some Selected Solutions to Technical Challenges} \label{sec_4}
\subsection{Basic Structure of the Proposed Platform}
The proposed platform has ultra-fast data performance modules ($>1$ GHz), multi-channel ultra-fast ($>1$ GSps) signal processing modules, and an ultrahigh bandwidth real-time signal processing application system with real-time application processor platform modules. It also has an operating system that allow easy implementation of a variety of applications.
For ultrahigh bandwidth signal processing, parallel processing is essential because the sampling rate of the input signal is faster than that of individual blocks. Hence, it is necessary to provide many logic and digital signal processor (DSP) units.
For real time operation, the proposed platform provides all functionalities including a hardware logic module, an real-time operating system-based module for real-time processing, and  universal operating system modules for the stand-alone execution/operation of various applications in one platform.
The proposed platform has host computer-based hardware/software operation and monitoring functions that enable rapid algorithm prototyping without any special knowledge of the hardware in the initial application development stage.
It also has a rich library of presentations to make it easier to perform various demos and tests.

\begin{figure}
\centering
\includegraphics[width=6.5in]{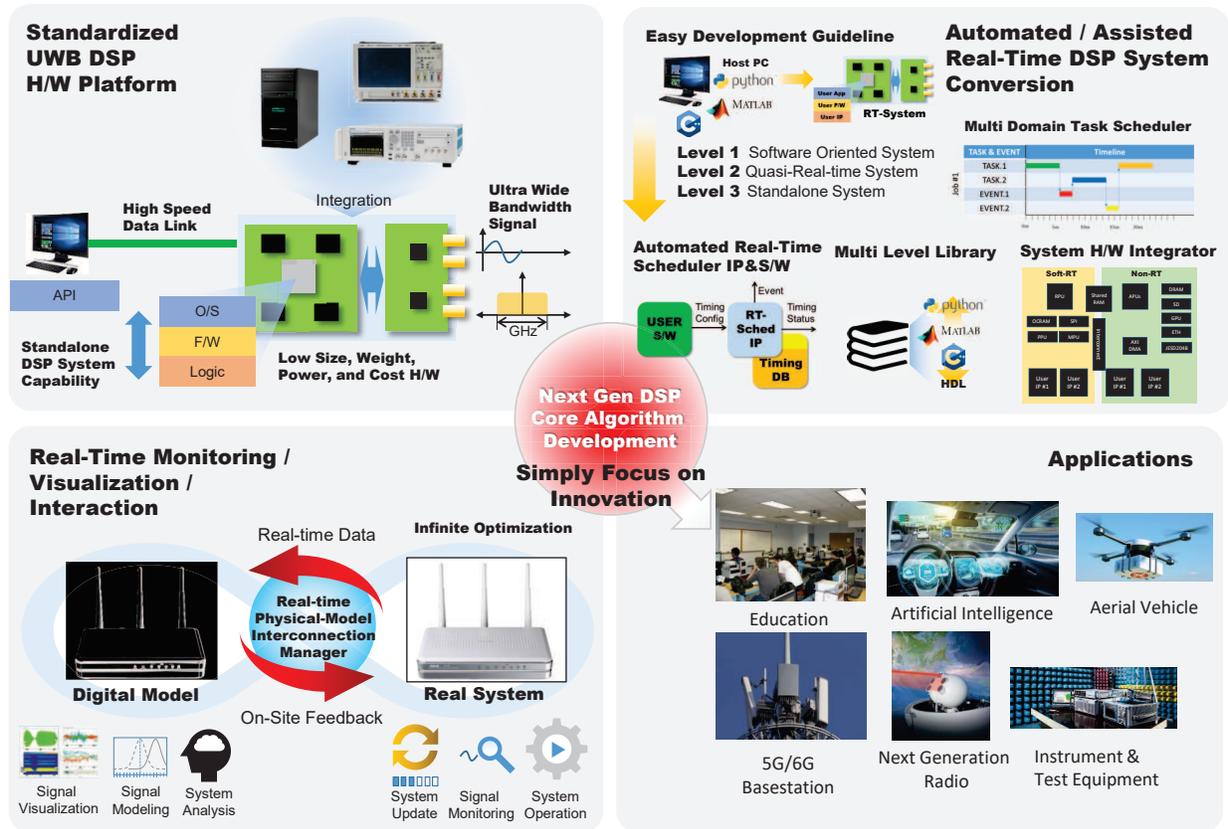}
\caption{Comparison of development process between conventional and proposed platforms.}
\label{Fig.2}
\end{figure}

\subsection{Key Elements for Ultrahigh Bandwidth SDR Platform}
The core elements of the proposed SDR platform are as the standardized ultrahigh bandwidth hardware platform, automated/assisted real-time DSP system conversion technology, and real-time monitoring/visualization/interaction technology. These elements are shown in Fig.~\ref{Fig.2}.
By applying this platform, users can focus on designing the core algorithms without worrying about the complex development processes inside the SDR platform. In addition, among the libraries verified through development, the library can be additionally expanded by recycling the unit function library possibly expanding the platform framework.

\subsubsection{Standardized Ultrahigh Bandwidth Hardware Platform}
The standardized ultrahigh bandwidth SDR hardware platform provides an ultrahigh speed connection function, a stand-alone operation function, and an ultrahigh bandwidth signal conversion function. The arbitrary waveform generator, digital signal analyzer, and digital signal processor functions can be implemented on a single board to achieve a small SWaP and low cost. In addition, the final product developed using the proposed platform can operate independently without a host computer.
The standardized hardware platform structure eliminates the unexpected problems that may occur in designing the system with various hardware and provides users with a platform that enables development in a relatively stable environment.

\subsubsection{Automated and Assisted Real-Time DSP System Conversion Technology}
The proposed platform provides users with a simple development and prototyping method for a real-time DSP system.
The final stand-alone system is established by the following three steps.
\\
\noindent 1) In step 1, a software-based simulation code for the target application is generated in the same way as in the traditional simulation code development process. In this step, the signal generation and analysis processes are not performed in real time, and the users perform the scheduling and reviews it temselves.
\\
\noindent 2) In step 2, the operation timing information of the code is additionally provided, and based on the provided timing table, the real-time scheduler located in the proposed platform manages the operation in real time. This allows the users to analyze and manage the behavior of the developed simulation code in real time. Finally, through an assisted optimization process, the PC-assisted system enables quasi-real-time operation.
\\
\noindent 3) In step 3, the operating code of the host computer is converted to that of  the proposed platform for stand-alone real-time operation. In this step, a multi-level library that can be used for programming languages such as MATLAB, Python, C++, and HDL is employed. The tasks provided through the multi-level library will automatically move the location from the host computer to the proposed platform. Thus, application systems that operate independently can be developed based on the proposed platform.

\subsubsection{Real-time Monitoring, Visualization, and Interaction Technology}
The proposed platform provides a monitoring and verification platform for the continuous testing and improvement of a developed system.
The developed system and its virtual model on a host computer are connected to synchronize the selected information.
The developed system continuously collects and processes real-time environment information needed to perform the tasks in real time and transmits it to the host computer through a high-speed communication interface. This information is conveniently visualized by emans of the graphical user interface of the host computer. The environment model of the user simulation is automatically updated based on this information allowing users to evaluate the performance of the developed system in the updated environment model that reflects the real-time environment. Based on this updated information in the host computer, the behavior of the developed system can be analyzed and optimized.
Thus, system development can be stably promoted through the optimization process, even in the development and operation stages.

\begin{figure}
\centering
\includegraphics[width=5.5in]{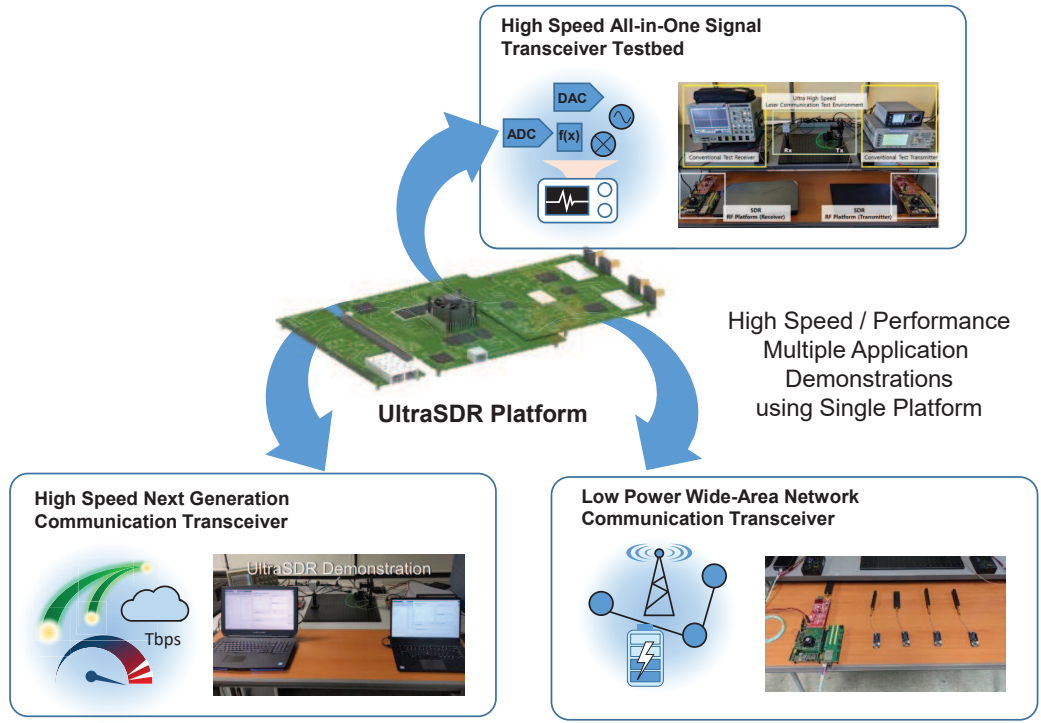}
\caption{Demonstration of applications on the example ultra-high bandwidth SDR platform.}
\label{Fig.4}
\end{figure}

\section{Some Selected Demonstration of Applications of The Proposed UltraSDR Platform} \label{sec_5}
In this section, we discuss some selected applications of the proposed UltraSDR platform in Fig.~\ref{Fig.4}.

\subsection{DCO-OFDM Demo}
As an advancement in the field of light-emitting diodes (LEDs), they have been developed to switch rapidly to different light intensity. This functionality has given rise to a novel communication technology known as visible light communication (VLC) wherein LED luminaries can be used for high-speed data transfer. Single-carrier modulation schemes have already been proposed recently; two typical types are on-off keying (OOK) and pulse-based modulation at high data rate. However, these schemes may be subjected to high inter-symbol interference (ISI) due to non-linear frequency response of VLC channels. To reduce ISI, direct current-biased optical orthogonal frequency division multiplexing (DCO-OFDM) has been proposed in the optical domain. In the DCO-OFDM system, all subcarriers are modulated by considering Hermitian symmetry to create real-valued output signals and a positive DC is added to make the signal unipolar.

\subsubsection{System Model}
The test-bed platform for the DCO-OFDM system demonstration is shown in Fig.~\ref{Fig.4}. 
A perfect synchronization is assumed because of the existence of digital control signal and host computers. We will explain the operation of the function block in the DCO-OFDM system programmed by MATLAB.
First, $M \cdot P$ data bits are generated where the $M$ is the chosen quadrature amplitude modulation (QAM) modulation order; the bits are modulated by the QAM encoder after the length is fixed such that it is divisible by $\log_2\left( M \right)$. After modulation, we get a QAM output signals of length $P$. The pilot signal is required to estimate channel characterization. Here, we can pre-run a traversing method to obtain the optimal pilot sequence that yields the minimum peak-to-average power ratio (PAPR) for different system parameter settings. Then, $N-P-1$ zeros (corresponding to DC value and $N = 0.5 N_{fft}$) are padded to the $P$ symbols and to convert the real-valued intensity modulated signal driving into an optical source, the complex conjugate of the mirror of the word of $N$ symbols is added to the latter before computing the $N_{fft}$-size inverse fast Fourier transform (IFFT). Moreover, a cyclic prefix is inserted to reduce ISI. Next, a DC offset is added, and the signals are transferred to real-positive values for optical transmission. Then, the modulated signals are transmitted by the SDR platform and distorted by the optical channel response. At the SDR receiver, the received signals are demodulated packet by packet. In the receiver part, cyclic prefix removal is done for the very first time because the DC value is concentrated in the null carrier after the FFT block which will be removed. Then, the signals are combined by two part: pilot and information sequence. The equalization model can estimate the channel response and then the QAM demodulator can generate the retrieved data bits.

\subsubsection{Discussions}
Among DCO-OFDM operations, FFT/IFFT operations require a large amount of computation, resulting in big bottlenecks when transmitting and receiving data. Therefore, to improve the system performance in the demonstration environment, the FFT/IFFT accelerator was applied to the transmitter and receiver enabling rapid conversion and restoration between the time-domain and frequency-domain signals. Thus, the conversion time was reduced by 2 ms per individual packet. 
In addition, the following operations were configured to be performed simultaneously to increase the transmission speed: i) data transfer from the host computer to the proposed SDR platform on the Tx side, ii) generation and transmission of the actual optical signal, iii) receiving of the actual optical signal at the proposed SDR platform on the Rx side, and iv) transferring the received optical signal to the host computer. 
As such operations  require longer time than actual data operation, the simultaneous, parallel operation enables users to perform these tasks twice as fast to achieve operation, and the operations close to real time.

\subsection{NRZ-OOK Demo}
In optical wireless communication systems, non-return-to-zero (NRZ) on-off-keying (OOK) modulations are widely used because of their easy implementation and cost effectiveness. NRZ-OOK modulation is a binary modulation in which ones are represented by one significant condition, usually a positive voltage, whereas zeros are represented by some other significant condition, usually a negative voltage, with no other neutral or rest condition.

\subsubsection{System Model}
The test-bed platform of the optical wireless NRZ system demonstration is shown in Fig.~\ref{Fig.4}.
In this system, we assume a simple noise-free configuration because the optical wireless communication system has a very high signal-to-noise ratio (SNR) system. In other words, we assume that the channel remains constant within the whole communication process because of the short distance between the transmitter and receiver in our system. The proposed platforms for the transmitter and receiver are connected to the host computer.
The packet synchronization is conducted by the host computer using external digital control interfaces and the detailed synchronization is conducted in the digital hardware synchronization part. Hence, the MATLAB designer can focus on designing the payload generation and reception.

\subsubsection{Function Description}
Since the proposed platform uses the linear digital to analog converter (DAC) and analog to  digital converter (ADC) for signal conversion, the system considers each sample as a 16-bit I/Q signal. The MATLAB program consists of two Tx and Rx functions so that various types of transmitting and receiving communication protocols can be designed based on this NRZ code program. These functions are designed to perform transmission/reception control by using a control signal between a transmitting device and a receiving device without the separate synchronization in MATLAB code. This synchronization part is designed in the hardware for reducing the burden on MATLAB designers.
The Tx function in a host computer receives the image/video data from storage devices or real-time devices and converts them into binary signals. This function directly communicates with the proposed platform through the Ethernet interface, and uninterrupted data transmission is possible by storing data in the buffer memory in the proposed platform.
The Rx function in the host computer is connected to the proposed platform through the Ethernet interface. When data are transmitted from the platform in the form of a Socket Client, this function receives the signals and reconstruct the image/video again.

\subsubsection{Discussions}
First, the image/video in the storage device is read, divided into packets, and transmitted using NRZ modulation. The packet was divided into the maximum transmission size of 64435 kB, and the data are delivered to the transmitting SDR platform. The transmitting SDR platform continuously generates data based on the received data and transmits the actual NRZ modulated optical signal. The SDR receiver converts the received signal and reconstructs it into packet data and decodes it into the image/video. Thus, data transmission of 200 MSPS was possible in the link stage, but real-time data transmission was not achieved because of the operation involved in the communication between the proposed platform and host computer. To solve this problem, the code developed in MATLAB should be converted to the hardware side.

\subsection{LoRa Transceiver Demo}
Several typical Internet of Thing (IoT) scenario involves power-limited devices that need to be connected to the Internet via wireless links. Against this background, low power wide area networks (LPWANs), which offer low data rate communication capabilities over ranges of several kilometers with limited energy, have emerged to complement the existing communication standards. Among those communication systems, the long range (LoRa) technique is one of the most promising LPWAN techniques and the LoRa Alliance has launched numerous IoT applications such as smart metering and smart grid.

\subsubsection{System Model}
LoRa employs chirp spread spectrum (CSS) modulation, which provides variable data rates by changing the spreading factor ($SF$). It is defined by its instantaneous frequency,
where $B$ is the bandwidth and $M$ is the number of possible waveforms at the output of the LoRa modulator given by $\left(M = 2^{SF} \right)$.
A discrete-time LoRa CSS modulated symbol with a spreading factor $SF$ of $6 - 12$ is usually standardized. 
Then the symbol duration is ${T_S} = {M \mathord{\left/ {\vphantom {M B}} \right.
 \kern-\nulldelimiterspace} B}$.

Ths LoRa Alliance \cite{LoRa5} also defined the PHY frame structure.
A frame is composed of a preamble,header, payload, and cyclic redundancy check (CRC). Note that the PHY-header and CRC are optional. The Preamble is constructed by $N_{pre}$ upchirps and 4.25 LoRa symbols (2 upchirps + 2 downchirps + 0.25 downchirps) as frame delimiters for synchronization.

A PHY-header contains the following frame information: a variable-length PHY-payload and CRC. More specifically, the transmission of an explicit header requires a spreading factor of at least 7, and the header is always transmitted with code rate $CR = 4$ at a reduced rate, the header also fits in an reduced interleaving matrix $\left(\in {(0,1)}^{(SF-2)\times(4+CR)}\right)$. Therefore, the header length must be equal to $SF-2$ codewords of 8 bits, i.e., 40 bits in total when $SF=7$. The header data thus have a length of 20 bits or 2.5 bytes. The left-to-right order of the PHY header has experimentally determined as follows: a single payload length in byte, followed by a nibble for the CR and the enabling state of MAC CRC, the high nibble (HN) of the header checksum, and finally the low nibble (LN) of the header checksum \cite{LoRa6}.

The test-bed platform of the  LoRa transceiver system demonstration is shown in Fig.~\ref{Fig.4}.
On the transmitter side, the input bits are first subjected to whitening. Then, Hamming coding, interleaving, and Gray indexing are  sequentially applied before modulation. LoRa uses CSS modulation for singling the preamble and data. The receiver performs synchronization and demodulation. Gray indexing, de-interleaving, de-whitening, and Hamming decoding are performed to recover the information data bits \cite{LoRa8}.

\subsubsection{Function Description}
In this part, we describe the functions and related algorithms of all models.

The CRC is an error-detecting code commonly used in digital networks and storage devices to detect accidental changes to raw data. It is optional for the payload but necessary for the header. 
There are two CRC-16 sequence for payload, CCITT and IBM, for paylload, and one CRC-8 sequence, CCITT, for the header.  
  The algorithm acts on the bits directly above the divisor in each step to finally generate a 16- or 8-bit CRC part.

DC-free data mechanisms are enabled to avoid DC bias caused by the payload, which may contain long sequences of 1's and 0's. A typical LoRa chip, e.g., SX1276, employs data whitening for the payload; this process is widely used for randomizing user data before radio transmission via an XOR operation of the payload and CRC checksum with a certain whitening pseudo random sequence. 

Hamming coding is an another method of error-correcting codes. Hamming codes can detect up to two-bit errors or correct one-bit errors without detecting uncorrected errors. Therefore, it is used in the LoRa to enhance the bit error rate performance at the expense of a certain rate of transmission $\left(4/\left(4+CR\right)\right)$. The LoRa system has five options $\left(CR =0,1,\cdots,4 \right)$ depending on the value of $CR$. The $\left(7,4\right)$ Hamming code can be used when $CR = 3$. Parity extended Hamming codes can be used when a higher error correction capability is desired, for example, (7,4) Hamming code with a parity bit can be used when $CR=4$. On the other hand, a plain parity code with rate $4/5$ or the identity are used here. Note that the structure of the Hamming code in the LoRa is not generally standardized.

Interleaving is another step used to decrease the symbol error rate. In the LoRa modulation, $\pm 1$ position demodulation errors have the highest probability. 
The LoRa interleaver is a diagonal interleaving solution that distribute the LoRa demodulation symbol error into different codewords and makes each codeword experience different bit reliability. The algorithm is showed as 

The LoRa specification also defines a ``reduced rate'' mode or ``low rate'' model. In this mode, $(SF-2)(CR+4)$ input bits are used to generate an $SF(4+CR)$ interleaving matrix. Therefore, in the reduced rate mode, a decreased data rate is traded for increased robustness to noise and fading. The PHY layer header of the LoRa frames is always transmitted in the reduced rate mode, whereas the payload bytes are only transmitted in reduced rate mode when $SF= 11$ or $SF= 12$.

Gray indexing is also widely used in digital communication system. However, the LoRa chip holds "reverse Gray" which still has the property that a mistaken in one adjacent symbol leads only to one bit error; hence, standard ``Gray'' coding should be performed to decode. Note that, in the reduced rate mode, the codewords always have two fixed zeros (or other 2 bits) in the LSB so that only one quarter Gray symbols are required. 

The FFT method is used to demodulate the LoRa signals. In this method, we first generate a base down-chirp signal $x^*$, which is the complex-conjugate of an upchirp. Next,  decimation is performed for in both down-chirp signal $x^*$ and received LoRa signal $y(n)$. Then dechirping is performed by finding the index  
$\max ({ DFT (x^* \cdot  y[n])})$.

As the first step in the demodulation process, the receiver must detect the LoRa preamble. The LoRa signal is very sensitive to synchronization and we first use a general cross-correlation method to find the starting point by moving the test window at a high speed.
Thereafter, we can determine the exact starting point and precisely move it to locate the starting point of the payload for demodulation.

\subsubsection{Discussions}
LoRa base station is designed to control the receiver device through the GUI located on the host computer side. The receiving device is designed to control the frequency and bandwidth through the configuration for LoRa Rx coding. The entire demonstration consists of a single LoRa base station and multiple LoRa nodes, and the LoRa base station is configured using the proposed platform. The LoRa node is configured to operate independently using the commercial MKR1310 LoRa device. Each LoRa node is individually equipped with a temperature and humidity sensor, a pyroelectric infrared detection sensor, and an ultrasonic distance sensor. Each sensor detects the surrounding environment at intervals (of $2 - 5$ seconds) periodically and transmits data according to the LoRa protocol. The proposed platform receives and displays the data on the GUI. Thus, it was confirmed that demodulation can be performed by effectively collecting data from multiple LoRa sensor devices.

\section{Conclusions} \label{conc}
In this paper, we proposed a programmable and reconfigurable stand-alone ultra-high bandwidth SDR platform without a host PC connection, expensive test equipment, or the need for skillful developers with sophisticated hardware/software expertise.
We first reviewed the current trends of the communication system technology and the their limitations in terms of implementation with the existing conventional SDR platform. Based on the review, we detailed the technical requirements and technical challenges faced by the ultrahigh bandwidth SDR platform.
Then, by considering the reviewed technical requirements and technical tasks, some selected solutions for the technical tasks were proposed. The demonstrations of a few applications of the proposed UltraSDR platform were discussed.

\bibliographystyle{ieeetran}
\bibliography{Ultra_SDR_Megazine}

\end{document}